\documentstyle[12pt]{article}
\newcommand{\be}{\begin{equation}}
\newcommand{\ee}{\end{equation}}
\newcommand{\bea}{\begin{eqnarray}}
\newcommand{\eea}{\end{eqnarray}}

\title{Irreducible Representations of an Algebra underlying Hidden
Symmetries of a class of Quasi Exactly Solvable Systems of Equations}

\author{Yves Brihaye  
\thanks{Department of Mathematical Physics,
University of Mons,
Place du Parc 20, B-7000 MONS, Belgium.}   \\
\and
Stefan Giller
\thanks {Department of Theoretical Physics,
University of Lodz,
Pomorska 149/153, 90-236 Lodz, Poland,
Work supported by Lodz University grants $n^o$
458 and $n^o$ KBN 2 P03B 076 10}  \\
\and
Piotr Kosinski$^{\ \dagger}$      \\ 
\and
Jean Nuyts$^{\ *}$        \\
}
\begin{document}

\begin{titlepage}
\maketitle
\thispagestyle{empty}
\begin{abstract}
The set of linear, differential operators 
preserving the vector space of couples of polynomials 
of degrees $n$ and $n-2$ in one real variable
leads to an abstract associative graded algebra ${\cal {A}}(2)$.
The irreducible, finite dimensional representations 
of this algebra are classified into five infinite discrete sets 
and one exceptional case. Their matrix elements are
given explicitely.
The results are related to the theory 
of quasi exactly solvable equations.
\end{abstract}
\vfill
\end{titlepage}

\section{Introduction}
The number of quantum mechanical problems which possess a complete
algebraic solution is rather limited. Some years ago 
\cite{turbiner1,ush,shifman}, 
several Hamiltonians
were exhibited which enjoy the property of having a finite number of
algebraic eigenvalues. That is to say that a part of their spectrum can be
obtained  by solving an algebraic (rather than a differential) eigenvalue
equation. Such equations and the corresponding linear operators are called
``Quasi Exactly Solvable'' (QES) (see \cite{turbiner2} for a recent review). 

In a one dimensional real space, the QES scalar operators
\cite{turbiner3}                                               
can be written
(after a suitable redefinition of the space variable and of the unknown
function) in the form of operators that preserve the vector space, say
$P(n)$, of polynomials $p_n(x)$ of maximal degree $n$ in the redefined space
variable. The
set of linear operators leaving $P(n)$ globally invariant coincides with
the envelopping algebra generated by the $spin$ $s=n/2$ representation of
the Lie group $SL(2,R)$ \cite{turbiner3}. This crucial observation allows a
classification of the QES operators and reveals the fact that they possess
a hidden symmetry. 

The notion of Quasi Exactly Solvable
systems of equations, fist addressed in \cite{shifman}, 
was extended recently in \cite{bk,bkgg,gonzalez}. 
In the case of two equations  in one real
variable , the relevant operators \cite{bk} (again after
suitable redefinitions) are those
preserving the vector space $P(m,n)$ of couples of polynomials
$p_m(x)$ and $p_n(x)$ of maximal degree $m$ and $n$ respectively in the
redefined variable. The question of identifying a hidden algebra behind
this QES system arises naturally. In \cite{bk}, 
it was shown that the set of
all linear operators preserving the space $P(m,n)$ coincides with the         
envelopping algebra of a representation of a 
particular graded algebra. The structure
of this algebra is strongly dependent on $\Delta=\mid m-n\mid$.  
Indeed, the natural composition law of the algebra is such that the
anticommutator of the ``fermionic'' generators is a polynomial
of degree $\Delta$ in the bosonic generators. Therefore, the
relevant algebra (which we denote ${\cal{A}}(\Delta)$)
is, in general, not of the Lie type. In the case
$\Delta=1$, it is ${\cal{A}}(1)$ and isomorphic 
to $osp(2,2)$.

Up to now, the abstract algebra ${\cal{A}}(\Delta)$ is obtained from a
particular representation~: the
representation given  by the linear operators preserving
$P(m,n)$. The structure constants of the algebra were obtained
by writing commutators and anticommutators among the generators
within this representation and imposing the Jacobi identities or
equivalently the associativity or the braiding relations.

The problem of finding all the irreducible representations
associated with the algebra of fixed $\Delta$ then arises
naturally. Apart from its direct algebraic interest, the
explicit construction of the representations also provides the
classification of all the QES systems (consisting of an
arbitrary number of equations) which possess the same underlying
symmetry as the system of two equations we started with. The
logical path is as follows. Once the abstract
algebra originating from a system of $2\times 2$ matrix 
operators preserving the space $P(m,n)$ has been obtained, one classifies 
all the
representations of the algebra and realizes them in terms of
matrix differential operators.

In this paper we reconsider and generalize the algebra ${\cal{A}}(2)$
and we classify all its irreducible representations.
It appears that the
abstract algebra has a rich set of irreducible representations
(in fact several inequivalent infinite families of them). 
Each of these representations can be
associated to a set of operators preserving a vector space
$P(n_1,n_2,\ldots,n_k)$ of $k$-tuple of polynomials in one real
variable and with maximal degrees $n_i$.

The paper is organized as follows. In section 2 we present the
algebra ${\cal{A}}(2)$. We show that it admits a generalisation
parametrized by three constants. We point out its symmetries,
its automorphisms and compute its two Casimir operators. In
section 3, we discuss the general properties of its
representations and present the tensorial operators relevant for
their construction. In section 4, we give the arguments leading
to the classification of the allowed representations. Section 5
then presents explicitely all the representations and in
particular the generic one. Finally section 6 indicates the way
to map the various representations in the formalism of QES equations.

\section{The abstract algebra}

\subsection{The algebra}

The algebra ${\cal{A}}(2)$ contains the
algebra $so(3)\otimes u(1)$ as a 4-generators Lie subalgebra
together with six more generators which behave as two vectors
under $so(3)$ and satisfy among themselves generalized
anticommutation relations. Let us define the algebra 
more precisely.

Denote by $T_i$ the three generators of the $so(3)$ Lie
subalgebra with the commutation relations
\be
\lbrack T_i,T_j\rbrack =f_{i,j}^{\phantom{i,j}k}\ T_k            
\label{comTT}
\ee

We work in a complex basis for $so(3)$ where the indices take
the values $+1,0$ and $-1$. In this basis the                             
$f_{i,j}^{\phantom{i,j}k}$ are
antisymmetric in $i$ and $j$ and zero except for 
\bea
f_{0,+1}^{\phantom{0,+1}+1}&=&1     \nonumber\\ 
f_{0,-1}^{\phantom{0,-1}-1}&=&-1     \nonumber\\ 
f_{-1,+1}^{\phantom{-1,+1}0}&=&1      
\label{fijk}
\eea
The corresponding metric $g_{i,j}$ 
\be
g_{i,j}={1\over 2} f_{i,k}^{\phantom{i,k}l} f_{j,l}^{\phantom{j,l}k} 
\label{gij1}
\ee
is symmetric and its non-zero elements are
\bea
g_{0,0}&=&g^{0,0}=1      \nonumber\\
g_{-1,+1}&=&g^{-1,+1}=-1     
\label{gij2}
\eea 
This metric is used to raise and lower the $so(3)$ indices
and allows us define
\bea
f_{i}^{\phantom{i}j,k}&=&g^{j,m}f_{i,m}^{\phantom{i,m}k} \nonumber\\
f^{i,j,k}&=&g^{i,m}g^{j,n}f_{m,n}^{\phantom{m,n}k} 
\label{fijkp}
\eea
Hence, the $so(3)$ Casimir operator $T^2$ is
\be
T^2=g^{i,j}T_iT_j=T_0^2-T_{-1}T_{+1}-T_{+1}T_{-1}
\label{T2}
\ee
and takes the eigenvalues $s(s+1)$ where $s$ is an integer or
a half-integer for selfadjoint representations.

The $u(1)$ operator $J$ commutes with the $T_i$
\be
\lbrack J,T_i\rbrack =0
\label{comJT}
\ee

The six extra generators, the $Q_i$ and the $\overline Q_i$,
vectors under $so(3)$
\bea
\lbrack  T_i,Q_j\rbrack &=&f_{i,j}^{\phantom{i,j}k}\ Q_k   \label{comTQ}\\
\lbrack  T_i,\overline{Q}_j\rbrack 
         &=&f_{i,j}^{\phantom{i,j}k}\ \overline{Q}_k  
\label{comTQbar}
\eea
have $J_{\rm{value}}=-1$ and $+1$ respectively i.e.
\bea
\lbrack J,Q_i\rbrack &=&-Q_i       \label{comJQ}\\
\lbrack J,\overline Q_i\rbrack &=&\overline Q_i  \label{comJQbar} 
\eea

The anticommutator of two $Q_i$ has $J_{\rm{value}}=-2$. Since there
exists in the algebra no generator with this $J_{\rm{value}}$, the 
only realistic possibility which fulfills 
the $so(3)\times u(1)$ invariance is
\be
\{Q_i,Q_j\}={2\over 3}g_{i,j}Q^2
\label{anticomQQ}
\ee
where the factor 2/3 is fixed by consistency and
\be
Q^2=g^{i,j}Q_iQ_j=3Q_0^2
\label{Q2}
\ee

Analogously the anticommutator of two $\overline Q_i$ is 
\be
\{ \overline Q_i,\overline Q_j\}={2\over 3}g_{i,j}\overline Q^2 
\label{anticomQbarQbar}
\ee
with
\be
\overline Q^2=g^{i,j}\overline Q_i\overline Q_j
       =3\overline Q_0^2   
\label{Qbar2}
\ee

For the anticommutator of a $Q_i$ with a $\overline Q_j$, which
has $J_{\rm{value}}$ zero, we write, a priori, the most general 
expression quadratic in
the operators $J$ and $T_i$ and with the correct $so(3)$ behaviour. 
We then impose the associativity
relations and find that there are three remaining free
parameters only
\bea
\{Q_i,\overline Q_j\}&=&\alpha
          \left ((T_iT_j+T_jT_i)-g_{i,j}(J^2+T^2)
          -2f_{i,j}^{\phantom{i,j}k}JT_k\right )      \nonumber\\
	 &&-\beta(g_{i,j}J+f_{i,j}^{\phantom{i,j}k}T_k)     \nonumber\\ 
	 &&-\gamma g_{i,j}                      
\label{anticomQQbar}
\eea

The full set of generalised commutation relations
is then given in (\ref{comTT}), (\ref{comJT}), 
(\ref{comTQ}), (\ref{comTQbar}), (\ref{comJQ}),
(\ref{comJQbar}), (\ref{anticomQQ}),
(\ref{anticomQbarQbar}), (\ref{anticomQQbar}). 
From now on, we will refer to this algebra as $\cal{A}$

If $\alpha$ is non-zero, as we will suppose henceforth, it can
be renormalised to any value by rescaling the operators $Q_i$ and/or 
$\overline Q_i$ by an appropriate multiplicative factor. 
In section 3, we have chosen to
normalise $\alpha$ to
\be
\alpha=\frac{1}{2}
\label{normalpha}
\ee

\subsection{Symmetries of the algebra $\cal{A}$}

In this section we present
the reparametrization which the free parameters
($\alpha,\beta,\gamma$) undergo when the operator $J$ is
subjected to a translation. Two automorphisms of the algebra
$\cal{A}$ are also given.

By a translation of $J$  
$$J\rightarrow J'=J-c$$ by a constant $c$, one obtains an
algebra equivalent to the original algebra through the
reparametrization
\bea
\alpha'&=&\alpha            \nonumber\\
\beta'&=&\beta+2\alpha c    \nonumber\\
\gamma'&=&\gamma+\alpha c^2 +\beta c   
\label{translation}
\eea
Obviously, by choosing $c$ suitably $\beta'$ or $\gamma'$ can be
made zero. Hence, there is
essentially only one free parameter in the algebra. 
The choice $\beta'=0$ is particularly interesting, as the
algebra with the constant term $\gamma'$ in the right hand side
can simply be interpreted as the
central extension of the algebra with both $\beta$ and $\gamma$ zero.

The two following quantities
\bea
I_1&=&2\alpha J+\beta               \label{I1}\\
I_2&=&\beta^2-4\alpha\gamma         \label{I2}
\eea    
are obviously invariant under the reparametrisation
(\ref{translation}) of the algebra. Note that the first quantity
is an operator. We will show that the discussion of the
representation can be best carried out in terms of the scalar invariant
$I_2$ and of the eigenvalues of the operator $I_1$.  

Another way of presenting the algebra is as follows. Using the
freedom in the definitions of the $Q_i$'s, the $\overline Q_i$'s
and the $J$, we have concluded that our algebra depends on one
significant parameter only (say $\gamma$ when $\beta=0$). As far as the
representations are concerned, we can write a single algebra
possessing the same set of irreducible representations as the
family of algebras parametrized by $\gamma$, at the expense of
introducing an additional generator $\Gamma$. Namely, we write  
\be
\{Q_i,\overline Q_j\}=\alpha\left(
               (T_iT_j+T_jT_i)-g_{i,j}(J^2+T^2)
                -2f_{i,j}^{\phantom{i,j}k} J T_k \right)
		 -\Gamma g_{i,j}
\label{antcomqqbar2}
\ee
and assume that $\Gamma$ commutes with all the generators.
Then, within irreducible representations, $\gamma$ becomes an
eigenvalue of $\Gamma$
\be
\Gamma\rightarrow \gamma I
\label{Gamma}
\ee
where $I$ is the identity matrix. 

The algebra which was obtained originally in \cite{bk} 
corresponds to the particular case
\bea
\alpha&=&1/2                 \nonumber\\
\beta&=&-1/2                  \nonumber\\
\gamma&=&0                      
\label{BK}
\eea

Let us also introduce two automorphisms of the algebra. 
\begin{enumerate} 
\item The interchange of $Q_i\leftrightarrow \overline Q_i$ 
made simultaneously with the replacement $J\rightarrow -J-\beta/\alpha$ is
an automorphism of the algebra. Precisely
\bea
Q_i&=&\overline Q_i'   \nonumber\\
\overline Q_j&=&Q_j'    \nonumber\\
J&=&-J'-\beta/\alpha   
\label{automorph1}
\eea
Under this automorphism (see (\ref{I1}))
\be
I_1=-I_1'
\label{automorph1p}
\ee
\item A second more trivial automorphism is given by the
multiplicative rescaling of $Q_i$ and $\overline Q_j$ by reciprocal
factors and more precisely
\bea
Q_i&=&\kappa Q_i'     \nonumber\\
\overline Q_j&=&{\overline Q_j'\over \kappa }  
\label{automorph2}
\eea
\end{enumerate}

\subsection{Casimir Operators of the algebra $\cal{A}$}

The two Casimir operators of the algebra $\cal{A}$ can also be computed. 
The
first one is of maximal fourth degree in the generators while
the second one is of sixth degree. 

The Casimir operators can be constructed with  
$so(3)$ invariant operators. Hence they can be constructed with
the operator $J$ and with scalars obtained from the vectors
$T_i$, $Q_i$ and $\overline Q_i$. We adopt the following
notation. Let $E_i$, $F_i$ and $G_i$ be arbitrary vectors. We
define in general
\bea
J_{EF}&=&g^{i,j}E_i  F_j   \nonumber\\
J_{EFG}&=&   f^{i,j,k}E_i F_{j} G_{k}   
\label{so3invariants}
\eea
When there are $\overline Q_i$ vectors
we have chosen, without loss of generality, to put them to the right. 
Among these invariants the two following identities are satisfied
\bea
3J_{TQQ} - 2 J_{TT} J_{QQ} + 6 J_{TQ} J_{TQ}&=&0   \nonumber\\
3 J_{T\overline Q\overline Q}-2 J_{TT} J_{\overline Q\overline Q}
+6 J_{T\overline Q} J_{T\overline Q}&=&0   
\label{identites}
\eea

Taking the most general form of the $so(3)\times u(1)$ invariant
polynomial of maximal degree six in the generators and imposing
commutation with the $Q$'s and the $\overline Q$'s one finds
that there are two independent Casimir operators of degree four
and six respectively. The Casimir operator of degree four $K_4$ is
\be
K_4 = R_0
      + R_1 J_{Q\overline Q}
      + R_2 J_{QQ}J_{\overline Q\overline Q} 
      + R_3 J_{TQ\overline Q} 
\label{Casgen4}
\ee
with
\bea
R_0&=&-\frac{\left(I_2-I_1^2\right) \left(I_2-(I_1-2\alpha)^2\right)}
           {64\alpha^2}    \nonumber \\  
&&+\frac{I_2+I_1^2-2\alpha I_1}{8} J_{TT}            
-{\alpha^2}{4} J_{TT}^2                   \\
R_1&=&\frac{I_1+2 \alpha}{2}             \\
R_2&=&\frac{1}{9}                       \\
R_3&=&\alpha                            
\eea
We have chosen to collect in $R_0$ all the terms which do not involve the
$Q$ or $\overline Q$ operators. It is amusing to note that if one acts on
a state of given $spin$ $s$, which means that $J_{TT}$ 
can be replaced by $s(s+1)$,
$R_0$ factorizes as
\be
R_0 =  - \frac{\left(I_2-(2 \alpha (s + 1) - I_1)^2\right)
               \left(I_2-(2 \alpha s + I_1)^2 \right)}
               {64 \alpha^2}
\ee  
Use of this property will be made later in the text. 

The Casimir operator of degree six $K_6$ (which is defined up to an
arbitrary contribution of the fourth order one) can conveniently be
chosen as
\bea
K_6&=&S_0                   
     + S_1 J_{Q\overline Q}  
     + S_2 J_{TQ\overline Q}           
     + S_3 J_{TQ}J_{T\overline Q} 
     + S_4 J_{Q\overline Q}^2     \nonumber\\   
     &&+ S_5 J_{QQ}J_{\overline Q\overline Q} 
     + S_6 (J_{QQ}J_{T\overline Q}^2
          +J_{TQ}^2J_{\overline Q\overline Q}) 
\eea
where we have grouped together in $S_i$ all operators which
don't depend on $Q$ or $\overline Q$                              
\bea
S_0 &=& 4 J_{TT} R_0                      \\
S_1 &=& \frac{52 \alpha^2 J_{TT} + 36 \alpha^2 
         + 16 \alpha J_{TT} I_1 + 12 \alpha I_1 - 3 I_2 +
            3 I_1^2}{4 \alpha}     \\ 
S_2 &=& \frac{12 \alpha^2 J_{TT} + 18 \alpha^2 
               + 6 \alpha I_1 - I_2 + I_1^2}{2 \alpha}   \\
S_3 &=& - 2 (  4 \alpha + I_1)      \\
S_4 &=& 1                               \\
S_5 &=& \frac{4 J_{TT} + 3}{3}           \\
S_6 &=&  -\frac{4}{3}
\eea
The fact that $S_0$ is proportional to $R_0$ is important and
will be used later.   

\section{The Representations}
In this section we want to describe the irreducible finite
dimensional representations of the algebra $\cal{A}$.
In the process of constructing explicitely the representations,
we have come across finite dimensional representations 
which are reducible without
being completely reducible. We have however not tried to
classify all such representations.

\subsection{The $J$,$s$ structure of the representations}

From the commutation (and anticommutation relations) of the
algebra $\cal{A}$ we infer the classification of the $J$,$s$
structure of the representations which is given in the theorem below.

Let us start by fixing some notations 
\begin{enumerate}
\item The eigenvalues $j$ of the operator $J$ within a
representation are discrete and range from $j_t$, the eigenvalue
with highest real part, to $j_b$, the eigenvalue with lowest real
part, in unit steps. 
\item We define the ``levels'' as the subspaces corresponding
to a given eigenvalue of $J$. The total number of levels $L$ is
thus given by
\be
L=j_t-j_b+1
\label{levelnumber}
\ee
\item The eigenspace corresponding to a given level
is the space of a finite (possibly
reducible) hermitian representation of the subalgebra $so(3)$.
It splits into a finite direct sum of spaces corresponding to a
certain set of $so(3)-spins$. Moreover its states can be
classified according to the eigenvalues of $T_0$. The general
basic states $\mid s,s_0,j>$ (which may have multiplicity higher
than one) thus satisfy 
\bea
T^2\mid s,s_0,j>&=&s(s+1)\mid s,s_0,j>      \label{state1}\\
J\mid s,s_0,j>&=&j\mid s,s_0,j>             \label{state2}\\
T_0\mid s,s_0,j>&=&s_0\mid s,s_0,j_t>       \label{state3}
\eea 
\end{enumerate}

\leftline{\bf{Classification Theorem}}

The complete set of finite dimensional irreducible
representations of the algebra $\cal{A}$ consists of five
main series and one exceptional case. 

\begin{description}
\item{G.} The generic series of representations
has $L$ levels (with $L \geq 4$, $s_t \geq 1$), acts on a
space of $4(L-2)(2s_t+1)$ dimensions and has the
following $J,s$ hierarchy
\be
\matrix{
J_{\rm{value}}&{\phantom{wide}}&{\rm{allowed}}\ s_{\rm{values}}      \cr   
j_t     &    &(s_t)                            \cr
j_t-1   &    &(s_t-1,s_t,s_t+1)                    \cr  
j_t-2   &    &(s_t-1,s_t,s_t,s_t+1)                  \cr  
\vdots  &    &\vdots                         \cr 
j_b+2   &    &(s_t-1,s_t,s_t,s_t-1)                  \cr  
j_b+1   &    &(s_t+1,s_t,s_t-1)                    \cr  
j_b     &    &(s_t)                            \cr
}
\label{hierarchyG}
\ee
\item{$H_+$.} The $H_+$ series
has $L$ levels (with $L \geq 2$), a
dimension $4(L-1)(s_t+1)-2$ and the
following $J,s$ hierarchy
\be
\matrix{
J_{\rm{value}}&{\phantom{wide}}&{\rm{allowed}}\ s_{\rm{values}}      \cr   
j_t     &    &(s_t)\phantom{,s_t+1}                            \cr
j_t-1   &    &(s_t,s_t+1)                   \cr  
\vdots  &    &\vdots                         \cr 
j_b+1   &    &(s_t,s_t+1)                   \cr  
j_b     &    &(s_t)\phantom{,s_t+1}                            \cr
}
\label{hierarchyH+}
\ee
When $s_t=s_b=0$ the $s=0$ states at levels $j_t-1$ and $j_b+1$
are absent and the dimension is decreased by $2$.
\item{$H_-$.} The $H_-$ series
has $L$ levels (with $L \geq 2$, $L$ even, $s_t \geq 1$), a
dimension $4(L-1)s_t+2$ and the
following $J,s$ hierarchy
\be
\matrix{
J_{\rm{value}}&{\phantom{wide}}&{\rm{allowed}}\ s_{\rm{values}}      \cr   
j_t     &    &\phantom{s_t-1,}(s_t)                \cr
j_t-1   &    &(s_t-1,s_t)                            \cr  
\vdots  &    &\vdots                             \cr 
j_b+2   &    &(s_t-1,s_t)                            \cr  
j_b+1   &    &(s_t-1,s_t)                            \cr  
j_b     &    &\phantom{s_t-1,}(s_t)                  \cr
}
\label{hierarchyH-}
\ee
\item{$T_+$.} The $T_+$ series
has $L$ levels (with $L \geq 2$, $L$ even, $s_t \geq 1/2$), a
dimension $4(L-1)(s_t+1)$ and the
following $J,s$ hierarchy
\be
\matrix{
J_{\rm{value}}&{\phantom{wide}}&{\rm{allowed}}\ s_{\rm{values}}      \cr   
j_t     &    &(s_t)\phantom{,s_t+1}                            \cr
j_t-1   &    &(s_t,s_t+1)                   \cr  
\vdots  &    &\vdots                         \cr 
j_b+1   &    &(s_t,s_t+1)                   \cr  
j_b     &    &\phantom{s_t,}(s_t+1)                            \cr
}
\label{hierarchyT+}
\ee
\item{$T_-$.} The $T_-$ series
has $L$ levels (with $L \geq 2$, $L$ even, $s_t\geq 3/2$), a
dimension $4(L-1)s_t$ and the
following $J,s$ hierarchy
\be
\matrix{
J_{\rm{value}}&{\phantom{wide}}&{\rm{allowed}}\ s_{\rm{values}}      \cr   
j_t     &    &\phantom{s_t-1,}(s_t)                            \cr
j_t-1   &    &(s_t-1,s_t)                   \cr  
\vdots  &    &\vdots                         \cr 
j_b+1   &    &(s_t-1,s_t)                   \cr  
j_b     &    &(s_t-1)\phantom{,s_t}                            \cr
}
\label{hierarchyT-}
\ee
\item{E.} The $E$ exceptional case
has $3$ levels, a dimension $4(2s_t+1)$, $s_t \geq 1$, and the
following $J,s$ hierarchy
\be
\matrix{
J_{\rm{value}}&{\phantom{wide}}&{\rm{allowed}}\ s_{\rm{values}}      \cr   
j_t       &    &(s_t)                            \cr
j_t-1     &    &(s_t-1,s_t+1)                   \cr  
j_t-2     &    &(s_t)                            \cr
}
\label{hierarchyE}
\ee
\end{description}

In the next section we discuss the arguments leading to a proof
of the classification theorem.

\subsection{General Properties of Representations}
 Let us now sketch the first part of 
the proof of the classification theorem.
This will be done in a few simple steps. 
The arguments will then be completed
in the following sections.
\begin{enumerate}
\item A finite dimensional representation of 
the algebra ${\cal{A}}$ provides obviously 
a finite dimensional representation (possibly reducible) 
of its $so(3)$ subalgebra. 
Since $so(3)$ is simple,
we conclude at once that the finite dimensional representations 
of ${\cal{A}}$ 
are direct sums of irreducible representations of $so(3)$, 
themselves equivalent to hermitian representations.
\item The operator $J$ is diagonalizable and its spectrum 
has the form~: 
\be
j_w=j_t-(w-1),\ w=1,\ \ldots\ ,L\ .
\label{spectrumJ}
\ee 
To see that this is the case, we first note that $J$, acting on 
a finite dimensional space, possesses at least one eigenvalue 
(real or complex) and its corresponding eigenspace. 
It follows from
the commutation rules (\ref{comJQ},\ref{comJQbar}) that the operators 
$\overline Q_i$ (resp $Q_i$) raise (resp. lower) the
$J$ eigenvalue by one. 
Consider the eigenspace of $J$ corresponding
to the eigenvalue $j_t$ with highest real part. 
This eigenspace (which we will call highest) is annihilated by
all $\overline Q_i$'s. Let us call normally 
ordered the product of operators in which
all the $\overline Q_i$'s stand rightmost. It follows 
from the commutation rules that any polynomial
in the generators can be normally ordered. 
Therefore, due to the irreducibility,
the whole representation space is obtained 
by acting on the highest eigenspace
(highest level) with all polynomials 
in the operators $J$, $T_i$ and $Q_i$ and of degree in the $Q_i$'s not
exceeding some non negative integer $L-1$. The diagonalizability 
of $J$ and the spectrum (\ref{spectrumJ}) follow from the last statement
and the commutation rules for the generators.
\item The eigenspace corresponding to the 
highest eigenvalue $j_t$ carries a representation of $so(3)$. 
It is easy to see that this must be an 
irreducible representation corresponding to a single $spin$ $s$,
conveniently labeled $s_t$. 
In order to show this, it is sufficient
to note that the space obtained by the 
action of all (normally ordered) polynomials in the generators 
(which do not decrease $J$) on the single 
$spin$ subspace of the highest level is an invariant subspace.
\item Let us call anti-normally ordered the 
product of generators in which all the
$Q_i$'s stand rightmost. 
The lowest level $j_b$ is annihilated by the $Q_i$'s and the whole
representation space is obtained 
by the action on the lowest space 
of (anti-normally ordered) polynomials in the generators. 
Therefore the lowest level carries also a
single $spin$ representation, say $s_b$, of $so(3)$.
\item  It is easy to see that the representation 
space is the linear span of all the
vectors obtained by the action of 
products of the $Q_i$ operators only (resp. the $\overline
Q_i$'s only) on all the states of
the highest (resp. lowest) level. 
In particular, 
the set of all monomials of degree $w$ (resp. $L-w-1$)
in $Q_i$'s (resp.  $\overline Q_i$'s) generates the level corresponding 
to the eigenvalue $j_t-w$.
\item The polynomials of degree $w$ in the $Q_i$ operators
can be classified according to
their $so(3)$ behaviour. 
Apart from the scalar $Q^2$ operator 
\be
Q^2\ \ \ \ \ \ \ \   \ J_{\rm{value}}=-2
\label{QQ2}
\ee 
which was defined
in (\ref{Q2}), it is useful to define a $so(3)$ vector (axial vector)
operator $A_i$ of $J_{\rm{value}}=-2$ and a scalar operator (pseudoscalar)
$P$ of $J_{\rm{value}}=-3$
\be
A_i=f_{i}^{\phantom{i}j,k}Q_jQ_k\ \ \ \ \ \ \ \ J_{\rm{value}}=-2
\label{axial}
\ee
\be
P=f^{i,j,k}Q_iQ_jQ_k\ \ \ \ \ \ \ \ J_{\rm{value}}=-3 .
\label{Pseudo}
\ee
It is not difficult to see that, as far as 
products of the $Q_i$ operators
are concerned, for even, say $J_{\rm{value}}=-2p,\ p\geq 1$, , there are
four independent operators
\bea
&(Q^2)^p     
&\ \ \ \ \ \ \ \ J_{\rm{value}}=-2p\ ,\ p\geq 1    
        \label{Q**even1}\\
&(Q^2)^{p-1} A_i    
&\ \ \ \ \ \ \ \ J_{\rm{value}}=-2p\ ,\ p\geq 1    
\label{Q**even2}
\eea
one being an $so(3)$ scalar and three others forming an $so(3)$ vector.
For odd, say $J_{\rm{value}}=-(2p+1),\ p \geq 1$, there are also
four independent operators
\bea
&(Q^2)^{p-1}P         
&\ \ \ \ \ \ \ \ J_{\rm{value}}=-(2p+1)\ ,\ p\geq 1    
    \label{Q**odd1}\\
&(Q^2)^{p} Q_i 
&\ \ \ \ \ \ \ \ J_{\rm{value}}=-(2p+1)\ ,\ p\geq 1    
 \label{Q**odd2}
\eea
Again there is an $so(3)$ scalar and an $so(3)$ vector.
The case of first order monomials is exceptional~: there is only one vector
operator
\be
Q_i\ \ \ \ \ \ \ \ \  J_{\rm{value}}=-1\ 
\label{Q**-1}
\ee
Obviously, analogous results are valid for monomials in the
$\overline Q_i$ operators.
\item Using the above classification and the $spin$-addition 
theorem, we conclude that
the following $spin$ structure emerges~: 
with $s_t$ the (unique) $spin$ corresponding 
to the level $j_t$, the level $j_t-1$ consists 
at most of the $spins$
$(s_t-1,s_t,s_t+1)$, while those corresponding 
to $j_t-w,\ w\geq 2$, consist at
most of the $spins$ $(s_t-1, s_t, s_t, s_t+1)$. Remark however
that if $s_t=0$ the only $spin$ which can be reached at level
$j_t-1$ is $1$ and only $0$ and $1$ at lower levels. If
$s_t=1/2$ the only $spins$ which can be reached are $1/2$ and $3/2$.
\item  Starting from the eigenspace corresponding 
to the lowest level with $J$ eigenvalue $j_b$ and
applying a
similar reasoning with $j_t$ replaced by $j_b$, 
$s_t$ by $s_b$ and $Q_i$ by
$\overline Q_i$ we conclude that the level 
$j_b+1$ consists at most of the $spins$
$(s_b-1, s_b, s_b+1)$, while the levels $j_b+w,\ w\geq 2$, 
consist at most of the
$spins$ $(s_b-1, s_b, s_b, s_b+1)$. Again $s_b=0$ or $s_b=1/2$
are special.
\item The four basic $Q$-monomials (of degree $w+1$) corresponding 
to the $J_{value}=-(w+1)$ can be obtained
from those (of degree $w$) corresponding to the $J_{value}=-w$ 
by a multiplication by the sole
operator $Q_0$.
Hence, the dimension of the space corresponding 
to the $j_t-w-1,\ w\geq 2$
level cannot be larger than the dimension of the space corresponding
to $j_t-w$ level.
 Following the same reasoning with $j_t$ replaced by $j_b$
and the $Q_i$'s replaced by $\overline Q_i$'s, 
we conclude that all levels corresponding to
$j_t-2\geq j\geq j_b+2$ have the same dimensionality.
 As $Q_0$ does not change the
$T_0$-content of the space the $spin$ content 
of all these levels is the same.
\item  It is obvious from the preceding discussion 
that $s_t$ and $s_b$ can
differ at most by one.
 Moreover, if they do differ, there can be only at
most two $spins$ in the intermediate levels.
 This is the case for the representations
$T_+$ (\ref{hierarchyT+}) and $T_-$ (\ref{hierarchyT-}).
\item If at any level $j$ the $spin$ structure 
is reduced to $(s_t+1,s_t-1)$ only, the next level 
(and the preceding level) has a
single $spin$ state $s_t$ only. Indeed the scalar (or pseudosalar)
which generates the states at level $j$ from the top level $j_t$
must give zero acting on the $\mid s_t,s_0,j_t>$ states. It
follows that the next vector operator (generating the states at
level $j-1$) is also zero. The states at level $j-1$ are reached
only by the scalar operator. Hence the conclusion. The same
holds for the preceding level by using the $\overline Q$ operators.
This shows that the exceptional case (\ref{hierarchyE}) is the
only one of its kind.
\item Finally, let us mention that we can also exclude, 
on general grounds,
the $spin$ patterns $(s_t-1,s_t,s_t)$ or 
$(s_t,s_t,s_t+1)$ or
$(s_t-1,s_t,s_t+1)$ for the levels 
$j_t-2\geq j\geq j_b+2$.
 We will only roughly sketch the arguments 
because the above
patterns are excluded by actual calculations 
in the following sections.
As it was stated above, all states 
of a given level are obtained by the
action of the operators (\ref{QQ2}-\ref{Q**-1}) 
on all the states of the $j_t$ level.
 Knowing what
is, a priori, the $spin$ content of the representation, 
we can construct out of
$T_i$'s the projection operators 
on the $spins$ $s_t-1$, $s_t$ and $s_t+1$.
Therefore we can construct explicitly 
all $spin$ representations.
Some algebra allows us then to show
that the two $spin$ $s_t$ representations 
at any level $j_t-2\geq j\geq j_b+2$ can be
chosen in such a way that the $Q_i$ operators 
do not mix the subspaces
corresponding to $spin$ $s_t-1$ and one of the subspaces 
with $spin$ $s_t$ 
with the subspaces corresponding to $spin$ $s_t+1$ and the second 
$spin$ $s_t$.
 It is then a matter of little effort
to show that no representation with three $spin$ subspaces on the levels
$j_t-2\geq j\geq j_b+2$ is allowed.
\end{enumerate}

\subsection{Eigenvalues of the Casimir operator}

Applied on the highest state $\mid s_t,s_0,j_t>$ the two Casimir
operators defined above take
on the values (remember our choice $\alpha=1/2$) 
\be
K_4=- \frac{1}{16}
(  I_2- (I_t+s_t)^2)\nonumber
( I_2-(I_t-(s_t+1)^2)
\label{Cas4val}
\ee
and
\be
K_6=4  s_t(s_t+1) K_4
\label{Cas6val}
\ee
where $I_t$ is the value taken by the 
invariant $I_1$ (\ref{I1}) on the highest state. 
\be
I_t= j_t+\beta
\ee
 
It is interesting to note that the ratio $K_6/4 K_4$ is simply
the value of the $so(3)$ Casimir for the highest state.

\vskip 0.5 true cm
\subsection{The Up, Down and Level tensorial operators.\hfill\break 
Identities}
Since the states allowed at every level $j_t-k$ ($0\leq k\leq q$)
have $spin$ $s_t-1$, $s_t$ and $s_t+1$ and since the $Q_i$ which
map these states on those of $J_{\rm{value}}=j_t-k-1$  can move the
$s_{\rm{value}}$ by at most 1 unit we are lead to define the following
three obviously relevant tensorial operators
\begin{enumerate}
\item The up operator $(U^{[s]}_i)_{m,n}$ which maps the
$n$th state $-s\leq n\leq s$ of the representations $s$ to the
$m$th state $-(s+1)\leq m\leq (s+1)$ of the representation $s+1$
and which is defined by 
\be 
(U^{[s]}_i)_{m,n}=(s+1)^{1\over2}(2s+3)^{1\over2}
C(s+1,m;1,i,s,n)
\label{Up}
\ee
where $C(j3,m3; j1,m1,j2,m2)$ is the Clebsh-Gordon coefficient
coupling $j1$ and $j2$ to make $j3$. (We use the Condon-Shortley
phase convention and normalisations.). Obviously this
coefficient is non-zero only if $m3=m1+m2$.
For every $i$, $U^{[s]}_i$ is a  $(2s+3)\times (2s+1)$ matrix.
\item The level operator $(L^{[s]}_i)_{n2,n1}$ which maps
the $n1$th state $-s\leq n1\leq s$ of the representations $s$ to
the $n2$th state $-s\leq n2\leq s$ of the representation $s$ and
which is defined by
\be
(L^{[s]}_i)_{n2,n1}=s^{1\over 2}(s+1)^{1\over 2}C(s,n2;1,i,s,n1)
\label{Level}
\ee
For every $i$, $L^{[s]}_i$ is a  $(2s+1)\times (2s+1)$ matrix .
\item The down operator $(D^{[s]}_i)_{n,m}$ which maps
the $m$th state $-(s+1)\leq m\leq (s+1)$ of the representations
$s+1$ to the $n$th state $-s\leq n\leq s$ of the representation
$s$ and which is defined by
\be
(D^{[s]}_i)_{n,m}=(s+1)^{1\over2}(2s+1)^{1\over 2}C(s,n;1,i,s+1,m)
\label{Down}
\ee
For every $i$, $D^{[s]}_i$ is a  $(2s+1)\times (2s+3)$ matrix.
\end{enumerate}
\vskip 0.5 true cm

These operators satisfy usefull identities which we now list and
which can be checked by explicit computations. 
In these identities we
have suppressed the obvious matrix indices.
\vskip 0.5 true cm
\begin{enumerate}
\item The down-up identity (for $s\geq 1$) is
\bea
D^{[s]}_iU^{[s]}_j&=&{1\over 2}(L^{[s]}_iL^{[s]}_j+L^{[s]}_jL^{[s]}_i)
	      +{2s+3\over 2}f_{i,j}^{\phantom{i,j}k}L^{[s]}_k
     \nonumber  \\
		&&-(s+1)^2g_{i,j}{\rm{I}}_{2s+1}
\label{Downup}
\eea
where ${\rm{I}}_{2s+1}$ is the unit matrix is the
$2s+1$-dimensional space of the representation $s$.
Extra identities are valid for $s=0$ and $s=1/2$ leading to
\be
D^{[0]}_iU^{[0]}_j=
		-g_{i,j}{\rm{I}}_{1}
\label{Downup1}
\ee
and
\be
D^{[\frac{1}{2}]}_iU^{[\frac{1}{2}]}_j=            
	      2f_{i,j}^{\phantom{i,j}k}L^{[\frac{1}{2}]}_k
		-2g_{i,j}{\rm{I}}_{2}
\label{Downup2}
\ee
\item The up-down identity (for $s\geq 0$) is
\bea
U^{[s]}_iD^{[s]}_j&=&
	 {1\over 2}(L^{[s+1]}_iL^{[s+1]}_j+L^{[s+1]}_jL^{[s+1]}_i)
	      -{2s+1\over 2}f_{i,j}^{\phantom{i,j}k}L^{[s+1]}_k
       \nonumber  \\
		&&-(s+1)^2g_{i,j}{\rm{I}}_{2s+3}
\label{Updown}
\eea
\item The non trivial level-level identity (for $s\geq 1/2$) is
\be
L^{[s]}_iL^{[s]}_j=
	 {1\over 2}(L^{[s]}_iL^{[s]}_j+L^{[s]}_jL^{[s]}_i)
	      +{1\over 2}f_{i,j}^{\phantom{i,j}k}L^{[s]}_k 
\label{Levellevel}
\ee
This is nothing else than the commutation relations of the
generators of $so(3)$. In other words, the factor in front of
the right-hand side in (\ref{Level}) has been chosen in such a way that
the $L_i$ satisfy exactly the commutation relations of the
abstract $T_i$ operators (\ref{comTT}). For $s=1/2$, the identity
can also be written more simply
\be
L^{[\frac{1}{2}]}_iL^{[\frac{1}{2}]}_j=
	      {1\over 2}f_{i,j}^{\phantom{i,j}k}L^{[\frac{1}{2}]}_k 
		+\frac{1}{4}g_{i,j}{\rm{I}}_{2}
\label{Levellevel1}
\ee
\item The up-up identity (for $s\geq 0$) is
\be
U^{[s+1]}_iU^{[s]}_j=U^{[s+1]}_jU^{[s]}_i
\label{Upup}
\ee
\item The down-down identity (for $s\geq 0$) is
\be
D^{[s]}_iD^{[s+1]}_j=D^{[s]}_jD^{[s+1]}_i
\label{Downdown}
\ee
\item The level-up identities are two~: namely (for $s\geq 1/2$)
\be
L^{[s+1]}_iU^{[s]}_j=-{1\over s}U^{[s]}_iL^{[s]}_j
		    +{s+1\over s}U^{[s]}_jL^{[s]}_i
\label{Levelup1}
\ee
and the relation (again for $s\geq 1/2$) 
which certifies that the up operator $U_i$
behaves as a $spin$ 1 operator
\be
L^{[s+1]}_iU^{[s]}_j=U^{[s]}_jL^{[s]}_i+f_{i,j}^{\phantom{i,j}k}U^{[s]}_k
\label{Levelup2}
\ee
For $s=0$, there is only one identity, namely
\be
L^{[1]}_iU^{[0]}_j=f_{i,j}^{\phantom{i,j}k}U^{[0]}_k
\label{Levelup3}
\ee
\item The level-down identities  are again two~: 
namely (for $s\geq 1/2$)
\be
L^{[s]}_iD^{[s]}_j={1\over s+2}D^{[s]}_iL^{[s+1]}_j
		    +{s+1\over s+2}D^{[s]}_jL^{[s+1]}_i
\label{Leveldown1}
\ee
and the relation (for $s\geq 1/2$)
which certifies that the down operator $D_i$
behaves as a $spin$ 1 operator
\be
L^{[s]}_iD^{[s]}_j=D^{[s]}_jL^{[s+1]}_i+f_{i,j}^{\phantom{i,j}k}D^{[s]}_k
\label{Leveldown2}
\ee
For $s=0$, we have one identity
\be
D^{[0]}_iL^{[1]}_j=f_{i,j}^{\phantom{i,j}k}D^{[0]}_k
\label{Leveldown3}
\ee
\end{enumerate}
\vskip 0.5 true cm
These are all the identities we need to try 
to construct the representations.

\section{The generic case}
\subsection{Form of the operators}
We present here the form of the different operators in the generic case, 
i.e. when the number of levels, say $L\equiv j_t-j_b+1$, is greater or
equal to four and all the states of the (\ref{hierarchyG}) are present.
\vskip 0.5 true cm
\begin{enumerate}
\item
Without loosing generality,
we can  assume that the operator 
$J$ can be diagonalized in blocks~:
\be
J=\left (
\matrix{
j_t\ {\rm{I}}_{(2s_t+1)} &0                   
&0                      &\cdots \cr
0              &(j_t-1)\ {\rm{I}}_{(6s_t+3)}  
&0                      &\cdots \cr
0              &0               
&(j_t-2)\ {\rm{I}}_{(8s_t+4)}         &\cdots \cr
\vdots       &\vdots                   
&\vdots                     &\ddots \cr
	}\right )
\label{J}
\ee
where the ``levels'' correspond to the subspaces of given 
$J$ (precisely to $J=j_t,j_{t}-1,\ \ldots\ ,j_b$) and where
${\rm{I}}_{m}$ is a diagonal unit matrix of dimension
$m$. The value of $m$ is $8s_t+4$ from the $J_{\rm{value}}=j_t-2$ 
down, except for the two last $J_{\rm{values}}$, namely for $j_b+1$
and for $j_b$ for which we have respectively $m=6s_t+3$
and  $m=2s_t+1$.
\item The $T_i$ operator assumes a block diagonal form made
of $L_i^{[s]}$ type matrices inside the diagonal blocks of given
$J_{\rm{value}}$. More precisely in the block $[j_t,j_t]$ one has
\be
T_i^{[j_t,j_t]}=L_i^{[s_t]}
\label{Tmat1}
\ee
in the block $[j_t-1,j_t-1]$
\be
T_i^{[j_t-1,j_t-1]}=\left(
		       \matrix{
	      L_i^{[s_t-1]}      & 0                &0          \cr
	      0              &L_i^{[s_t]}             &0          \cr
	      0              &0                 &L_i^{[s_t+1]}  \cr  
			     } 
	       \right )
\label{Tmat2}
\ee 
Then starting from the block $[j_t-2,j_t-2]$ one has                 
\be
T_i^{[j_t-p,j_t-p]}=\left(
			\matrix{
       L_i^{[s_t-1]}    & 0             &0          &0              \cr
       0              &L_i^{[s_t]}      &0          &0              \cr
       0              &0              &L_i^{[s_t]}  &0              \cr
       0              &0              &0          &L_i^{[s_t+1]}    \cr  
			       } 
	       \right )
\label{Tmat3}
\ee 
where $2 \leq p \leq j_b+2$.
For the two last blocks, namely $[j_b+1,j_b+1]$ and
$[j_b,j_b]$, the content can be inferred from (\ref{hierarchyG})
$ T_i^{[j_b+1,j_b+1]} \equiv T_i^{[j_t-1,j_t-1]}$ and 
$T_i^{[j_b,j_b]}\equiv T_i^{[j_t,j_t]}$.

\item The operators $Q_i$ have their representations in
terms of the blocks situated exactly one step below the
diagonal blocks (\ref{J}). 
In the block $[j_t-1,j_t]$ one has a $(6s_t+3)\times(2s_t+1)$
matrix of the form
\be
Q_i^{[j_t-1,j_t]} \equiv Q_i^{[1]} =
              \left (
			\matrix{ 
	   c_{11}^{[1]} D_i^{[s_t-1]}          \cr
	   c_{21}^{[1]} L_i^{[s_t  ]}          \cr
	   c_{31}^{[1]} U_i^{[s_t  ]}        \cr
			      }
	       \right )
\label{Qmat1}
\ee
where $c^{[1]}_{11},c^{[1]}_{21}$ and $c^{[1]}_{31}$ 
are three arbitrary constants.
The $[j_t-2,j_t-1]$ block is a $(8s_t+4)\times(6s_t+3)$ matrix which is written
\be
Q_i^{[j_t-2,j_t-1]}  \equiv Q_i^{[2]}             
	 =    \left (
		       \matrix{ 
 c_{11}^{[2]} L_i^{[s_t-1]}  &c_{12}^{[2]} D_i^{[s_t-1]}     &0   \cr
 c_{21}^{[2]} U_i^{[s_t-1]}    &c_{22}^{[2]} L_i^{[s_t]}  
		    &c_{23}^{[2]} D_i^{[s_t]}  \cr
 c_{31}^{[2]} U_i^{[s_t-1]}    &c_{32}^{[2]} L_i^{[s_t]}  
		   &c_{33}^{[2]} D_i^{[s_t]}  \cr
 0                                    &c_{42}^{[2]} U_i^{[s_t]}
		   &c_{43}^{[2]} L_i^{[s_t+1]}  \cr
			   }
	     \right 
			       )
\label{Qmat2}
\ee
where there are ten arbitrary constants. It is to be remarked
that the elements $c^{[2]}_{13}$ and $c^{[2]}_{41}$ are zero since,
 by elementary
properties of tensorial products in $so(3)$, there is no
operator of $spin$ 1 connecting the space of $spin$ $s-1$ to the
space of $spin$ $s+1$.  The block $[j_t-3,j_t-2]$, a
$(8s_t+4)\times(8s_t+4)$ matrix has the form
\be
Q_i^{[j_t-3,j_t-2]}     \equiv Q_i^{[3]}=  
	  \left (
			\matrix{ 
 c_{11}^{[3]} L_i^{[s_t-1]} &c_{12}^{[3]} D_i^{[s_t-1]}  
	  &c_{13}^{[3]} D_i^{[s_t-1]}       &0                  \cr
 c_{21}^{[3]} U_i^{[s_t-1]}     &c_{22}^{[3]} L_i^{[s_t]}  
	  &c_{23}^{[3]} L_i^{[s_t]}   &c_{24}^{[3]} D_i^{[s_t]}  \cr
 c_{31}^{[3]} U_i^{[s_t-1]}     &c_{32}^{[3]} L_i^{[s_t]}   
	  &c_{33}^{[3]} L_i^{[s_t]}   &c_{34}^{[3]} D_i^{[s_t]}  \cr
 0                      &c_{42}^{[3]} U_i^{[s_t]} 
 &c_{43}^{[3]} U_i^{[s_t]} &c_{44}^{[3]} L_i^{[s_t+1]}  
			       }
	       \right )
\label{Qmat3}
\ee
This depends on 14  coefficients $c^{[3]}$.
The following blocks have the same structure except the two last
blocks which can be read off (\ref{hierarchyG}). Namely
a $(6s_t+3)\times(8s_t+4)$ matrix 
\bea
&&Q_i^{[j_b+1,j_b+2]} \equiv Q_i^{[L-2]} \nonumber   \\
            &&\quad =
	  \left (
			\matrix{ 
 c_{11}^{[L-2]} L_i^{[s_t-1]} &c_{12}^{[L-2]} D_i^{[s_t-1]}  
	  &c_{13}^{[L-2]} D_i^{[s_t-1]}       &0                       \cr
 c_{21}^{[L-2]} U_i^{[s_t-1]}     &c_{22}^{[L-2]} L_i^{[s_t]}  
	  &c_{23}^{[L-2]} L_i^{[s_t]}   &c_{24}^{[L-2]} D_i^{[s_t]}  \cr
 0                      &c_{32}^{[L-2]} U_i^{[s_t]} 
 &c_{33}^{[L-2]} U_i^{[s_t]} &c_{34}^{[L-2]} L_i^{[s_t+1]}  
			       }
	       \right )\nonumber\\
\label{Qmat4}
\eea
and a $(2s_t+1)\times(6s_t+3)$ matrix 
\be
Q_i^{[j_b,j_b+1]}  \equiv Q_i^{[L-1]}=         
	  \left (
			\matrix{ 
 c_{11}^{[L-1]} U_i^{[s_t-1]}\ \     & c_{12}^{[L-1]} L_i^{[s_t]} \ \   
	   &c_{13}^{[L-1]} D_i^{[s_t]}  \cr
			       }
	       \right )
\label{Qmat5}
\ee

\item The form of the operators $\overline Q_i$ is obviously
analogous to that of the $Q_i$ but the blocks are situated one
step above the diagonal blocks i.e. in the positions $[j_t-p,j_t-p-1]$,
again in terms of the tensor operators $U_i^{[s]}$, $D_i^{[s]}$ and
$L_i^{[s]}$,  and the corresponding constants are labelled $\overline
c$. More precisely, the specific form of the operators 
$\overline Q_i$ can be obtained from the form of the $Q_i$ above by
transposition and the interchange of the tensor operators $D_i$
and $U_i$ for the same $s{\rm{value}}$, with no change on the $L_i$. 
\end{enumerate}
 
For later convenience, it is useful to define the following 
matrices with the  coefficients $c$ and $\overline c$~:
\be
 C^{[1]} =      \left (
			\matrix{ 
	   c_{11}^{[1]}           \cr
	   c_{21}^{[1]}           \cr
	   c_{31}^{[1]}         \cr
			      }
	       \right ) \ \ , \ \ 
\overline C^{[1]} =      \left (
			\matrix{ 
\overline c_{11}^{[1]} &\overline c_{12}^{[1]} &\overline c_{13}^{[1]} \cr
			      }
	       \right )
\label{CCBAR}
\ee
and so on for $C^{[2]}$ (4$\times$3 matrix),
 $\overline C^{[2]}$ (3$\times$4 matrix) \dots,  i.e. the matrices
obtained from the $Q$'s and the $\overline Q$'s by replacing the operators
$U,D$ and $L$ by the number one.

\subsection{The equations}

With the forms of the operators given above in (\ref{J},\ref{Qmat5}) the
equations (\ref{comTT},\ref{comJT},\ref{comTQ},\ref{comTQbar},
\ref{comJQ},\ref{comJQbar}) are automatically fullfilled. What
remains to be imposed are the anticommutation relations of the $Q_i$
and $\overline Q_i$ among themselves (\ref{anticomQQ},
\ref{anticomQbarQbar},\ref{anticomQQbar}).
Obviously these relations are used to determine
the parameters $c$ and $\overline c$. 
Using the identities of section (3.4), it appears that (\ref{anticomQQ}) 
results in the following constraints
on the matrices $C^{[k]}$
\be
C^{[k+1]}\ C^{[k]} = 0\quad , \quad k = 1, \cdots, L-2
\label{orthoc}
\ee
Similarly, using (\ref{anticomQbarQbar}) together with the identities, 
we obtain
\be
\overline C^{[k]}\ \overline C^{[k+1]} = 0 \quad , \quad k=1,\cdots, L-2
\label{orthocbar}
\ee
The equations on $c, \overline c$  obtained by imposing the relations
(\ref{anticomQQbar})  cannot be written is such a compact way. 
We observe that the 
anticommutator of the left hand side take a block diagonal form~:
\bea
\lbrace Q_a, \overline Q_b\rbrace^{j_t,j_t} &=&  
{\phantom {Q^{[p]}_a \overline Q^{[p]}_b +\ \ \ \ \ }}
\overline Q^{[1]}_b Q_a^{[1]} \label{diag1}\\
\lbrace Q_a, \overline Q_b\rbrace^{j_t-p,j_t-p} &=&
Q^{[p]}_a \overline Q^{[p]}_b\ \  +\ \  \overline Q^{[p+1]}_b Q^{[p+1]}_a
  \ , \ {\rm {for}}\  1 \leq p \leq L-2 \label{diag2} \\
\lbrace Q_a, \overline Q_b\rbrace^{j_b,j_b} &=&  
 Q^{[L-1]}_a \overline Q_b^{[L-1]} \label{diag3}
\label{diagqqbar}
\eea

Within each block, the identities of section (3.4) can be used to 
put the expressions as combinations of linearly independent operators.
The identification 
of the coefficients of the independent operators of (\ref{diagqqbar})
with those of the right hand side of (\ref{anticomQQbar})
then leads to a system of equations
for products of parameters $c$ with parameters $\overline c$.

\subsection{Similarity transformations}

It is not difficult to see that there remains some freedom 
in the definition
of the operators $Q$ and $\overline Q$. This is related 
to the fact that we can
rescale independently the vectors in the different $spin$ 
representations and
mix in an arbitrary way the two $spin$ $s$ representations 
within one level.
This freedom results in the following redefinition of 
the matrices $C^{[k]}$
and $\overline C^{[k]}$ 
\be
C'^{[k]} =U_{k+1} C^{[k]} U^{-1}_k
\label{sim1}
\ee
\be
\overline C'^{[k]} =U_{k} C^{[k]} U^{-1}_{k+1}
\label{sim2}
\ee
for $k=1, \cdots, L-1$. Here $U_1$ and $U_L$ are (non zero) numbers,
$U_2$ and $U_{L-1}$ are $3\times 3$ diagonal invertible               
matrices and $U_k$
are invertible matrices of the form
\be
U_k = \left(\begin{array}{cccc}
\mu_k & 0 & 0 &0\\
0 &\nu_k &\lambda_k &0\\
0 &\theta_k &\rho_k &0\\
0 & 0 & 0 & \sigma_k
\end{array}\right)\quad , \quad k=3, \cdots, L-2.
\label{sim3}
\ee
All parameters $\mu_k, \nu_k, \ldots$ 
 appearing in $U_1,\ldots, U_L$ are complex numbers. We shall use
this freedom to put the matrices $C^{[k]}$ in a particularly simple form.

\subsection{Canonical form of the $C$ matrices}

We now determine the parameters $c$ and $\overline c$ in the case
when $L\geq 4$ (lower dimensional cases are treated later)
and assuming that the $Q$ and $\overline Q$ operators connect
all, a priori possible, pairs of $spins$ between consecutive levels
(i.e. with the pattern of (\ref{hierarchyG})).

We first concentrate on equations (\ref{orthoc}). 
These equations, together with the
similarity transformations freedom 
(\ref{sim1},\ref{sim2},\ref{sim3}) 
allows us to determine all 
the matrices $C$ in function of only one
parameter. The canonical forms of them read as follows
\be
C^{[1]} = \left(\begin{array}{c}
1\\
1\\
1
\end{array}\right) \ \ \ , \ \ \ 
C^{[2]} = \left(\begin{array}{ccc}
1 &-1 &0\\
1 &-1 &0\\
0 &-1 &1\\
0 &-1 &1
\end{array}\right)
\label{canonc1}
\ee
\be
C^{[3]} = C^{[4]} = \cdots = C^{[L-3]} =
\left(\begin{array}{cccc}
1 &-1 &0&0\\
1 &-1 &0&0\\
0 &0 &-1&1\\
0 &0 &-1 &1
\end{array}\right)
\label{canonc2}
\ee
\be
C^{[L-2]} = \left(\begin{array}{cccc}
1 &-1 &0&0\\
1 &-1 &-X&X\\
0 &0 &-1&1
\end{array}\right) \ \ \ , \ \ \ 
C^{[L-1]} = (1, -1, X)
\label{canonc3}
\ee
This parametrisation greatly simplifies the solution 
of the other equations.
In particular it leads, for (\ref{diagqqbar}), 
to linear constraints in the $\overline c^{[k]}$'s.
Moreover, the form of eq.(\ref{diagqqbar}) allows one to solve
these linear equations recursively in $[k]$.

Let us now discuss how these equations are solved.
 Imposing the relation for the first block leads to a
self consistent linear system for the parameter 
$\overline c^{[1]}_{1k} (k=1,2,3)$.
The solution of this system reads 
\bea
 \overline c^{[1]}_{11} &\equiv A &= {(I_t-s_t-1)^2-I_2\over{2s_t(2s_t+1)}} 
                    \label{A}   \\
\overline c^{[1]}_{12} &\equiv B &=
{s_t(s_t+1)-(I_t-1)^2+I_2\over{2s_t(s_t+1)}}  \label{B}   \\
\overline c^{[1]}_{13} &\equiv C &=
{(I_t+s_t)^2-I_2\over{2(2s_t+1)(s_t+1)}}     \label{C}
\eea
where we define $I_t \equiv j_t+\beta$, i.e. the value of 
the operator $I_1$
for $J=j_t$ (remember we have normalized $\alpha = 1/2$).
Remark that the Casimir value $K_4$ is nothing else but  $-AC/16$.
If we hadn't normalized $C^{[1]}$ to unit values (\ref{canonc1}),
the unique solution for $A$, $B$ and $C$ would have corresponded in
general to $A=\overline c^{[1]}_{11} c^{[1]}_{11}$,
$B=\overline c^{[1]}_{12} c^{[1]}_{21}$
an to $C=\overline c^{[1]}_{13} c^{[1]}_{31}$. Hence the
the restriction of the eigenspace of $J_{\rm{value}}=j_t-1$ to a
space with two $so(3)$ $spins$ instead of three leads to the
vanishing of one of the functions $A,B$ or $C$.

It is usefull to note that under the involution 
$\left\{ s_t\leftrightarrow -(s_t+1) \right\}$,
\bea
B&=&B   \ \Biggr\vert_{s_t\rightarrow -(s_t+1)}\quad \nonumber \\
C&=&A\ \Biggr\vert_{s_t\rightarrow -(s_t+1)}
\label{ABCsymmetry}
\eea
which means that $B$ is invariant while $A$ and $C$ are interchanged.
Analogous involutions will occur at higher levels.

\par Considering the second block in (\ref{diagqqbar}) 
we obtain an (apparently)
overdetermined system of 19 linear equations in 10 variables. The solution
nevertheless exists, is unique and reads
\be
\overline C^{[2]} =\left(\begin{array}{cccc}
{s_t+1\over s_t}B &C-{1\over s_t}B &-C &0\\
-{s_t+1\over s_t}A &{1\over s_t}A &-{1\over{s_t+1}}C &-{s_t\over{s_t+1}}C\\
0 &-A &A+{1\over{s_t+1}}B &{s_t\over{s_t+1}}B
\end{array}\right)
\ee
Under the involution $\left\{ s_t\leftrightarrow -(1+s_t) \right\}$ 
(see (\ref{ABCsymmetry})) 
the  elements of $\overline C^{[2]}$ are
interchanged as follows  
$\overline C^{[2]}_{i,j}\leftrightarrow \overline C^{[2]}_{4-i,5-j}$.  

For the next blocks, the number of equations and of 
variables are respectively
24 and 14. Again the equations are compatible with each other
and provide a unique solution.

The structure of the matrices $C^{[k]}$ ($k>2$) suggests to use
a similar block decomposition for the $\overline C^{[k]}$, i.e. 

\be
\overline C^{[k]} =\left(\begin{array}{cc}
\overline C_{11}^{[k]} &\overline C_{12}^{[k]}\\
\overline C_{21}^{[k]} &\overline C_{22}^{[k]}
\end{array}\right)
\label{Cbarblockform}
\ee
The solution for $\overline C^{[3]}$ reads

\be
\overline C_{11}^{[3]} ={1\over{s_t^2}} \left(\begin{array}{cc}
-A(s_t+1)-B+Cs_t &A+B(1-s_t)+Cs_t(s_t-1)\\
-A(s_t+1)^2-B(s_t+1) &A(s_t+1)-B(s_t^2-1)
\end{array}\right)
\label{c311}
\ee
\bea
\overline C_{22}^{[3]} &=&{1\over{(s_t+1)^2}}\times\nonumber\\ 
&&\times\left(\begin{array}{cc}
-Bs_t(s_t+2)-Cs_t &s_tB-s_t^2C\\
A(s_t+1)(s_t+2)+B(s_t+2)+C &-A(s_t+1)-B+Cs_t
\end{array}\right)\nonumber\\ 
\label{c322}
\eea
\be
\overline C_{12}^{[3]} = -C {s_t\over{(s_t+1)^2}} \left(\begin{array}{cc}
s_t+1 &0\\
1 &s_t
\end{array}\right)
\label{cbar312}
\ee
\be
\overline C_{21}^{[3]} = -A {(s_t+1)\over{s_t^2}} \left(\begin{array}{cc}
s_t+1 &-1\\
0 &s_t
\end{array}\right)
\label{cbar321}
\ee
We remark that the blocks $\overline C_{12}$ and
$\overline C_{21}$ are respectively proportional to the matrix
element $\overline c^{[1]}_{13}$  and  $\overline c^{[1]}_{11}$.
Moreover, it is easy to check that $\overline C^{[3]}_{21}$ and
$\overline C^{[3]}_{12}$ are related by the involution
$\left\{ s_t\leftrightarrow -(1+s_t) \right\}$ 
\be
     \overline C^{[3]}_{21} 
    = \sigma_1\overline C^{[3]}_{12}\sigma_1
\ \Biggr\vert_{s_t\rightarrow -(s_t+1)}
\label{map1}
\ee
and that
\be
    \overline C^{[3]}_{22} 
    =\sigma_1  \overline C^{[3]}_{11} \sigma_1 
\ \Biggr\vert_{s_t\rightarrow -(s_t+1)}
\label{map2}
\ee
where $\sigma_1$ is the first Pauli matrix.

\par The conditions coming from the next blocks 
determine a set of recursive
relations for the elements of the 4$\times$4 matrices 
$\overline C^{[k]}$,$[k]>3$. 
The structure of the matrices $\overline C^{[k]}$ is such that the
equations relative to the four 2$\times$2 blocks defined above decouple. 

First, the blocks $\overline C^{[k]}_{12}$ and  $\overline C^{[k]}_{21}$
satisfy the recurrence relations
\be
\overline C^{[k+1]}_{12} ={s_t\over{s_t+1}} \overline C^{[k]}_{12}
\ee
Hence, for all $[k] \geq 3$ 
\be
\overline C_{12}^{[k]} = -C {
s_t^{k-2} \over (s_t+1)^{k-1}
} 
\left(\begin{array}{cc}
s_t+1 &0\\
1 &s_t
\end{array}\right)
\ee

The matrices $\overline C^{[k]}_{21}$ are obtained analogously to 
eq.(\ref{map1}) by 
\be
     \overline C^{[k]}_{21} 
   =  \sigma_1\overline C^{[k]}_{12}\sigma_1
\ \Biggr\vert_{s_t\rightarrow -(s_t+1)}
\label{map3}
\ee
The recursive equations for the block $\overline C^{[k]}_{11}$ are
\be 
\overline C^{[k+1]}_{11} + \overline C^{[k]}_{11} =
k^2 M_2 + k M_1 + M_0
\label{recursion}
\ee
where 
\be
M_2 = {1\over{2 s_t^2}} \left(\begin{array}{cc}
-1 &1\\
-1 &1
\end{array}\right)
\label{recursion2}
\ee
\be
M_1 =  {1\over{s_t^2}} \left(\begin{array}{cc}
I_t      &-(s_t+I_t)\\
-(s_t-I_t) &-I_t
\end{array}\right)
\label{recursion1}
\ee
\be
M_0 = {1\over{2 s_t^2}} \left(\begin{array}{cc}
s_t(s_t+1)+I_2 - I_t^2               & (s_t+I_t)^2 - I_2 \\
-\left((s_t-I_t)^2 - I_2 \right)   & -\left( s_t(s_t-1)+I_2-I_t^2 \right)
\end{array}\right)
\label{recursion0}
\ee
The solution, with the appropriate boundary condition
$\overline C^{[3]}_{11}$ (\ref{c311}),  can easily be obtained
\bea 
\overline C^{[k]}_{11} =&-&(-1)^k \overline C^{[3]}_{11}
+ (k^2 + 9(-1)^k) {M_2 \over 2}
+ (k + 3(-1)^k) {M_1-M_2 \over 2}\nonumber\\
&+& (1 + (-1)^k) {2M_0 - M_1 \over 4}
\label{solutionck}
\eea

The recursion equation for the matrix $\overline C^{[k]}_{22}$ 
is completely analogous to the equation for $\overline C^{[k]}_{11}$.
The solution is simply obtained, for $k\geq 3$ by the involution
$\left\{ s_t\leftrightarrow -(1+s_t) \right\}$
\be
     \overline C^{[k]}_{22} 
   = \sigma_1\overline C^{[k]}_{11}\sigma_1
\ \Biggr\vert_{s_t\rightarrow -(s_t+1)}
\label{map22}
\ee
If we define the matrices $N_i$ from the matrices $M_i$ by the
involution
\be
     N_i 
   = \sigma_1 M_i \sigma_1
\ \Biggr\vert_{s_t\rightarrow -(s_t+1)}
\label{mapMN}
\ee
the $\overline C^{[k]}_{22}$ satisfy an equation of the form 
(\ref{recursion}) with the $M_i$ replaced by the $N_i$

In this approach, we have solved (\ref{diagqqbar}) 
starting from the highest value $j_t$ of $J$ and going down. 
Alternatively, these equations 
 can be solved by starting from the lowest value $j_b$ of $J$
and proceeding by going up. This procedure gives for instance
the following values for $\overline C^{[L-1]}$
\bea
\overline C^{[L-1]}_{11} &\equiv \tilde A &=
 {(I_b+s_t+1)^2-I_2\over{2s_t(2s_t+1)}} \\
\overline C^{[L-1]}_{21} &\equiv \tilde B &= 
 {( I_b+1)^2-I_2-s_t(s_t+1)\over{2s_t(s_t+1)}} \\
\overline C^{[L-1]}_{31} &\equiv \tilde C &= 
 {( I_b-s_t)^2-I_2\over{2(s_t+1)(2s_t+1)X}}
\eea
with 
\be
  I_b =  j_b+\beta = I_t-L+1
\ee
One condition for the representation to be irreducible is that
$\tilde A \tilde B \tilde C \neq 0$. The special values of the
parameters which
annihilate $\tilde A \tilde B \tilde C$ are discussed in the
next section.
It should also be remarked that, due to the presence of the, 
yet undetermined, parameter $X$ in eq.(\ref{canonc3}), 
the involution, analogous to (\ref{ABCsymmetry}),  
has to be changed slightly to     
\bea
\tilde B&=&\tilde B   \ \Biggr\vert_{s_t\rightarrow -(s_t+1)}\quad \\
\tilde C&=
  &\frac{1}{X}\left(\tilde A\ \Biggr\vert_{s_t\rightarrow -(s_t+1)}\right)
\label{ABCtildesymmetry}
\eea

For $\overline C^{[L-2]}$ one obtains 
\be
\overline C^{[L-2]} =\left(\begin{array}{ccc}
-{s_t+1\over s_t}\tilde B &{s_t+1\over s_t}\tilde A  &0                     \\
-{1\over s_t}\tilde B - X\tilde C &{1\over s_t}\tilde A &X \tilde A    \\
\tilde C  &-{1\over s_t+1}\tilde C &- \tilde A +{1\over s_t+1} \tilde B \\
0  &{s_t\over{s_t+1}}\tilde C &-{s_t\over{s_t+1}}\tilde B
\end{array}\right)
\ee
and, using the block form (\ref{Cbarblockform}) for $\overline C^{[L-3]}$ 
\be
\overline C_{11}^{[L-3]} ={1\over{s_t^2}} \left(\begin{array}{cc}
-\tilde A(s_t+1)+\tilde B-\tilde CXs_t 
        &(s_t+1)(\tilde A(s_t+1)-\tilde B) \\
-\tilde A+(s_t-1)(-\tilde B+\tilde CXs_t) 
        &(s_t+1)(\tilde A+\tilde B(s_t-1))
\end{array}\right)
\label{cL311}
\ee
\be
\overline C_{22}^{[L-3]} ={1\over{(s_t+1)^2}} \left(\begin{array}{cc}
\tilde Bs_t(s_t+2)+\tilde CXs_t 
&(s_t+2)(-\tilde A(s_t+1) +\tilde B) + \tilde CX\\
s_t(\tilde B- \tilde CXs_t)   
      &-\tilde A(s_t+1)+\tilde B-\tilde CXs_t
\end{array}\right)
\label{cL322}
\ee
\be
\overline C_{12}^{[L-3]} 
       = - \tilde AX {s_t+1 \over{s_t^2}} \left(\begin{array}{cc}
s_t+1 &0\\
1 &s_t
\end{array}\right)
\label{cbarL312}
\ee
\be
\overline C_{21}^{[L-3]} 
      = - \tilde C{s_t \over{(s_t+1)^2}} \left(\begin{array}{cc}
s_t+1 &-1\\
0 &s_t
\end{array}\right)
\label{cbarL321}
\ee
The matrices $\overline C^{[L-k]}$ (for $k>3$)  can then be determined
recursively.

Remarkably, the value of the matrix 
$\overline C^{[L-3]}$ predicted from the
recurrence relations, starting from the top, 
match with the value 
obtained by solving the equation
from below provided only one relation among 
the parameters $j_t, L, s_t$ is imposed.
The value of the parameter $X$ (see (\ref{canonc3})) 
is also uniquely predicted by
this procedure. The expressions of the constraint and of $X$ appear to be
quite different according to the parity of $L$. 
\begin{enumerate}
\item For $L$ even, $X$ is uniquely determined to be 
\be
X=- ({s_t\over{s_t+1}})^{L-2} {(2s_t+L-1)^2-4I_2\over{(2s_t-L+3)^2-4I_2}}
\label{Xeven}
\ee
At the same time, the consistency of all equations fixes 
uniquely $I_t$ as a function of $L$~: 
\be
I_t =  \frac{L-1}{2}  
\label{consistencyeven}   
\ee
As a consequence, the spectrum of the operator $I_1$ is
$I_t, I_t-1,\ldots,I_b=-I_t$,  symmetric around zero. 
The spectrum of $J$ is, obviously, the
spectrum of $I_1$ shifted by $-\beta$. In (\ref{Xeven}) we have
to exclude the limiting cases $X=0$ and $X=\infty$
\be
X=0 \rightarrow I_2=\frac{(2 s_t+L-1)^2}{4} 
\ee
which corresponds to a special limit (\ref{H+even}), or
\be 
X=\infty \rightarrow I_2=\frac{(2 s_t-L+3)^2}{4}
\ee
which corresponds to (\ref{H-even}).
\item For $L$ odd, the parameter $X$ is fixed as   
\be
X = - ({s_t\over{s_t+1}})^{L-2} {2s_t+L-1\over{2s_t-L+3}}
\label{Xodd}
\ee
while the quantity $I_t$ is determined by the equation
\be
    (I_t - \frac{L-1}{2})^2 = I_2 +\frac{(2s_t+L-1)(2s_t-L+3)}{4}
\label{consistencyodd}   
\ee
which allows for two values of $I_t$ and, hence, of the spectrum of $I_1$.
The two corresponding representations transform into each 
other under the automorphism (\ref{automorph1}). 
In (\ref{Xodd}) the limiting value $X=\infty$ has to be excluded, i.e.
\be
X=\infty \rightarrow L=2s_t+3 
\ee
These values correspond to the special case (\ref{H+odd}). 
\end{enumerate}

Hence, for fixed values of $L$ and of $s_t$, 
all the matrix elements of $C^{[k]}$ and of $\overline C^{[k]}$
are uniquely determined.
 
We further checked that the equations (\ref{anticomQbarQbar}) 
(which leads to quadratic
equations among the $\overline C^{[k]}$) are automatically obeyed.

The discussion above demonstrates that the algebra ${\cal{A}}$ 
admits an infinite tower of irreducible representations 
labelled by the integers $L$ and $2s_t$. 
Their dimensions  $d=4(2s_t+1)(L-2)$ can be arbitrarily large
and the spectrum of the operator $J$ is quantized.
We refer them to as to the generic representations.

This result contrasts in many respects with its counterpart 
for the graded Lie algebra $osp(2,2)$. In this case,
the generic irreducible and finite dimensional 
representations consist of three levels only,
with the following $spin$ content \cite{ritt} 
\be
 (s_t),(s_t-1,s_t+1),(s_t)
\ee
analogous to the exceptional case (\ref{hierarchyE}).

\subsection{Special limits\hfill\break
            The $H_+$ and $H_-$ series of representations }

We will now discuss the way to obtain the representations
of type $H_+$ and $H_-$ of the theorem. We have constructed all the matrix
elements of these representations
by solving all the equations restricted, at the
start, to the relevant eigenspaces of given $J$ (see 
(\ref{hierarchyH+}), (\ref{hierarchyH-})).
We have then realized  that all
the necessary information can be extracted from the generic 
representation extensively described in the previous section. 

Let us start with the representation of type $H_+$. In this case,
the equations corresponding to the bloc (\ref{diag1}) 
lead to a same system of three equations
in two variables
(for instance $B=\overline c^{[1]}_{12}c^{[1]}_{21}$ 
and $C=\overline c^{[1]}_{13}c^{[1]}_{31}$).
These equations are obviously
identical to those determining $A,B$ and $C$ 
(\ref{A},\ref{B},\ref{C})
when $A$ is put to zero.
Since the solution in the generic case was unique,
the new equations are compatible with each under the condition
that 
the missing variable $A=\overline c^{[1]}_{11}c^{[1]}_{11}$ vanishes.
Similarly, the equations associated to the last bloc (15)
leads to the condition $\tilde A = 0$ (see (44)).
Therefore, two necessary conditions for a representation of type $H_+$
to occur read
\bea
A&=0 \quad      \longrightarrow \quad  I_2 &= (I_t-s_t-1)^2    \nonumber\\ 
\tilde A&=0 \quad\longrightarrow \quad I_2 &= (I_b+s_t+1)^2  
\label{conditionH+}
\eea
In this case, many elements of the matrices $\overline C^{[k]}$ 
($3\leq k \leq L-3$) (in particular $\overline C^{[k]}_{12}$ and
$\overline C^{[k]}_{21}$)
vanish and the generic representation can consistently be
restricted to the subspace
\be
(s_t), (s_t,s_t+1), (s_t,s_t+1),\ \cdots_t\ , (s_t,s_t+1),(s_t) \ \ \ 
                   , \ \ L \ {\rm {levels}}  
\ee
The restriction of the matrices $C^{[k]}$ and $\overline C^{[k]}$
to the lower-right 2$\times$2 block provides the relevant  
matrix elements. The matrices $\overline C^{[k]}_{22}$ obey the
recurence relation of the generic case but the initial condition if fixed 
already by  $\overline C^{[2]}_{22}$ (see (\ref{mapMN})) i.e.
\bea 
\overline C^{[k]}_{22} &=& (-1)^k \overline C^{[2]}_{22}
+ (k^2 - 4(-1)^k) {N_2 \over 2}
+ (k - 2(-1)^k) {N_1-N_2 \over 2}    \nonumber   \\
&&+ (1 - (-1)^k) {2N_0 - N_1 \over 4}
\label{solutionckh}
\eea
with
\be
\overline C^{[2]}_{22} =\frac{1 }{s+1} \left(\begin{array}{cc}
 -C & -sC \\
 B  &  sB
\end{array}\right)
\ee

We have checked that all the commutation relations are then
satisfied provided that the consistency relations (\ref{consistencyeven})
(for $L$ even) or (\ref{consistencyodd}) (for $L$ odd) are also satisfied.

Summarising the results, we conclude that
\begin{enumerate} 
\item The representation $H_+$ exists, when the number of levels
$L$ is {\bf{even}}, if 
\bea
   I_t &=& {L-1 \over 2} \nonumber \\
   I_2 &=& {1\over 4} (L-2s_t -3)^2 
\label{H+even}
\eea
Since $L$ and $2s_t$ are integers, 
the parameter $I_2\equiv\beta^2-2\gamma$ is
restricted to a discrete set of special values.
\item The representation $H_+$ exists, when the number of levels
$L$ is {\bf{odd}}, if 
\bea
   L  &=&2s_t+3 \nonumber    \\ 
   I_2 &=& (I_t-s_t-1)^2 
\label{H+odd}
\eea
and we see in particular that $s_t$ has to be restricted to be an integer.

Though the identities among the tensorial operators, 
which were written in section 3.4, take a
different form when $s=0$ or $s=1/2$, about all the cases which
involve these $spins$ in one of the spaces behave in a normal
way with one important exception pertaining to the
representation $H_+$ when $s_t=0$. Indeed, the $J,s$
hierarchy reduces to
\be
(0),\ (1),\ (0,1),\ \ldots\ ,\ (0,1),\ (1),\ (0) 
\label{szeroH}
\ee
The $spin$ $s=0$ space is missing at levels $j_t-1$ and $j_b+1$.
The equations which have to be satisfied are less numerous and
we have obtained only one restriction instead of two both for
even and odd $L$.
\begin{description}
\item{a.} For $L$ even, the restriction is
\be
I_t=\frac{L-1}{2}\ \ \ {\rm{for}}\ L\ {\rm{even\ and}}\ s_t=0
\ee
Moreover, for example, the matrix $\overline C^{[1]}$ has 
only one entry which should be non zero. Other matrix elements
$\overline C^{[3]}(1,2),\ldots$
cannot be zero also. 
This excludes some
values for $I_2$. Precisely 
\be
I_2\neq \frac{(L-(2p-1))^2}{4}+(p-1)(p-2)\ \ \ 
{\rm{for}}\ \ p=1,2,\ldots,[\frac{L}{4}]+1
\ee
\item{b.} For $L$ odd, the restrition reads
\be
I_2=I_t^2-(L-1)I_t+\frac{(L-1)(L-2)}{2}
\ \ \ {\rm{for}}\ L\ {\rm{odd\ and}}\ s_t=0
\ee
Again the matrix element of $\overline C^{[1]}$ and some other
matrix elements have to be non zero. This excludes some
values for $I_t$. Precisely
\be
I_t\neq (2p+1)/2 \ \ \ {\rm{for}}\ p=1,2,\ldots,L-3
\ee
\end{description}

Let us now discuss, in the same way, the conditions of
occurrence of the representation 
$H_-$. Following the same reasoning as above, one shows that the necessary
conditions for this representations to exist are $C=\tilde C =0$, i.e.
\bea
C&=0 \quad\longrightarrow \quad I_2 &= (I_t+s_t)^2 \nonumber \\
\tilde C&=0\quad \longrightarrow \quad I_2 &= (I_b-s_t)^2
\label{conditionH-}
\eea
Then, the generic representation can consistently be
restricted to the subspace
\be
(s_t), (s_t-1,s_t), (s_t-1,s_t),\ \cdots\ , (s_t-1,s_t),(s_t) \ \ \ 
                   , \ \ L \ {\rm {levels}}  
\ee

Compatibility of the equations (\ref{conditionH-}) with the
consistency relations (\ref{consistencyeven})
(for $L$ even) or (\ref{consistencyodd}) (for $L$ odd) lead to
the conclusion.
\item The representation $H_-$ exists only when the number of
levels $L$ is {\bf{even}} and if 
\bea
   I_t &=& {L-1 \over 2} \nonumber    \\
   I_2 &=& {1\over 4} (L+2s_t -1)^2  
\label{H-even}
\eea
The parameter $I_2$
is again restricted to a discrete set of special values.
\end{enumerate}

\subsection{Special limit\hfill\break
            The exceptional representation}

The exceptional representation can be obtained as a special limit of the
generic series of representations. In complete analogy with the arguments
given in the preceding section, it is obtained by putting to zero the
parameter $B=\overline c^{[1]}_{12} c^{[1]}_{21}$ (\ref{B})
\be
(I_t-1)^2=I_2+s_t(s_t+1)
\label{Exceptional}
\ee 
which means that the $spin$ states at level $j_t-1$ have to be restricted
to the values $s_t-1$ and $s_t+1$ and 
thus that the states corresponding to $spin$
$s_t$, at that level, have to be discarded. 
It is then easy to see that the representation closes by the addition of
the next and last level $j_t-2=j_b$ containing one set of $s_t$ states only. 

Given $s_t$ there are two $j_t$ fulfilling (\ref{Exceptional}). These
two cases are related by the first automorphism of the algebra
(\ref{automorph1},\ref{automorph1p}) which transforms $I_t$ into $-I_b$
and thus $I_t-1$ into $-I_b-1=-(I_t-1)$. 

In obvious notation, the representation is completely determined by 
\be
 C^{[1]} =      \left (
			\matrix{ 
	   1         \cr
	   1           \cr
			      }
	       \right ) \ \ , \ \ 
C^{[2]} =      \left (
			\matrix{ 
1 & -1 \cr
			      }
	       \right )
\label{Cexeptional}
\ee
\be
\overline C^{[1]} ={\frac{1}{2s_t+1}}
                 \left (
			\matrix{ 
 -(I_t-s_t-\frac{3}{2})\  &\ (I_t+s_t-\frac{1}{2})  \cr
			      }
	       \right )
\ee
and
\be 
\overline C^{[2]} ={\frac{1}{2s_t+1}}
                 \left (
			\matrix{ 
	  I_t+s_t-\frac{1}{2}           \cr
	  I_t-s_t-\frac{3}{2}         \cr
			      }
	       \right )
\label{Cbarexeptional}
\ee

\subsection{Gluing of representations\hfill\break
            The $T_+$ and $T_-$ representations}

We have also constructed explicitely the representations
corresponding to the cases
$T_+$ of the classification (\ref{hierarchyT+}). 
The case $T_-$ (\ref{hierarchyT-}) 
can be obtained from the case $T_+$ by the
automorphism (\ref{automorph1}) of the algebra.
\begin{description}
\item{a.} 
The representation $T_+$ exists only if $L$,
the number of levels, is {\bf{even}}
and if the invariants $I_1$ and $I_2$ are fixed as follows~:
\bea
I_t &=& s_t+ {L+1 \over 2} \nonumber  \\ 
I_2 &=& {(L-1)^2 \over 4} 
\label{T+even}
\eea
Note the restricted values of the $I_2$ invariant.
The representation $T_+$, however, does not exist if  
$s_t=0,s_b=1$. 
\item{b.} 
The representation $T_-$ exists only if $L$,
the number of levels, is {\bf{even}}
and if the invariants $I_1$ and $I_2$ are fixed as follows~:
\bea
I_t &=& -s_t+ {L-1 \over 2} \nonumber  \\ 
I_2 &=& {(L-1)^2 \over 4} 
\label{T-even}
\eea
Note the restricted values of the $I_2$ invariant.
The representation $T_-$, however, does not exist if  
$s_t=1,s_b=0$. 
\end{description}

Again these
representations are strongly related to the generic representations.
In order to perceive the connection let us first remark
that the representations $H_-$ (when the levels $j_t$ and $j_b$ have
$spin$ $s$) and $H_+$ (when the levels $j_t$ and $j_b$ have $spin$ $s-1$)
have the same $spin$ pattern as far as their ``internal" part
is concerned~:
\bea
&&H_-(j_t\ {\rm{with}}\ spin\ s)     \nonumber\\
         &&\quad\quad\quad (s),  (s-1,s),
           \  \ldots_t\ ,  (s-1,s), (s) \\
&&H_+(j_t\ {\rm{with}}\ spin\ s-1)    \nonumber\\
         &&\quad\quad\quad (s-1),(s-1,s),
           \  \ldots\ , (s-1,s), (s-1)
\eea
It is therefore tempting to try to match
the upper part of the first of these representations 
with the lower part of the second one to produce 
a $T_-$ ($j_t$ with $spin$ $s_t=s$, $j_b$ with $spin$ $s_b=s_t-1$) 
representation. 
The alternative matching 
would produce an $T_+$ ($j_t$ with $spin$ $s_t=s-1$, $j_b$ with $spin$
$s_b=s_t+1=s$) representation.

For the representation $H_-$ (upper part of $T_-$) 
the relevant part of the operators $Q_i$ 
is given by the matrix (see (\ref{canonc2}))
\be
C^{[k]}_{11} = \left(\begin{array}{cc}
1 &-1\\
1 &-1\end{array}\right)
\ee
and the relevant part of the operator $\overline Q_i$ 
is parametrized by $\overline C^{[k]}_{11}$ (see (\ref{solutionck})).
For the representation $H_+$ (lower part of $T_-$) the relevant matrices  
are respectively
\be
C^{[k]}_{22} = \left(\begin{array}{cc}
-1 &1\\
-1 &1\end{array}\right)
\ee
and $\overline C^{[k]}_{22}$ (see (\ref{map22})) with however $s_t$
replaced by $s_t-1$.
We see at once that a smooth matching requires to change
the sign of the operators $Q_i$ and $\overline Q_i$ of one
of the two representations involved (using e.g. 
the automorphism (\ref{automorph2}) with $\kappa =-1$).
Imposing the equality between the 
blocks  $\overline C^{[k]}_{11}$ 
appearing in the $H_-$ representation with the $-\overline C^{[k]}_{22}$
block appearing in the $H_+$ representation (with, remember, $s_t$
shifted into $s_t-1$)  
at any level $k$ implies the conditions (\ref{T+even}) for the
representation $T_+$ 
can be seen as coming from $A(s_t)=0$ and 
$\tilde C(s_t\rightarrow s_t+1)=0$. In an analogous way, the
conditions (\ref{T-even}) for the representation $T_-$ 
come from $C(s_t)=0$ and 
$\tilde A(s_t\rightarrow s_t-1)=0$.  

\section{Relations with QES operators}

All the operators constructed above can be represented by
linear differential operators preserving some vector space, 
say P$(n_1,n_2,\ldots ,n_k)$,
whose vectors are  
\be
    p_{n_1}(x) ,\  p_{n_2}(x),\  \ldots\  ,\ p_{n_k}(x)
\ee
(where $p_{n}(x)$ are polynomials of degree at most $n$ in $x$) 
for suitable values of $k$ and $n_k$. 

For $N=1$, this can be achieved by means
of the following correspondence ($n\equiv 2s$)
\bea
  L_a^{[s]} \longrightarrow 
& J_a(n) \equiv &
\left( {d \over dx} ,\ \ x{d \over dx} 
      - {n \over 2} ,\ \ x^2 {d \over dx} - nx \right) 
\nonumber\\
U_a^{[s]} \longrightarrow 
       & q_a  \equiv&  \left( 1,\ \ x,\ \  x^2 \right) 
\nonumber\\
  D_a^{[s]} \longrightarrow 
       & \overline  q_a(n) \equiv&
\left( {d^2 \over dx^2} ,\ \  (x{d \over dx} - n - 1){d \over dx},\right.
\nonumber    \\
&& \left.(x{d \over dx} - n-1)(x{d \over dx} -n-2)\right ) 
\label{caseN3}
\eea
The operators $J_a(n)$ are the ones introduced by Turbiner
\cite{turbiner1}.
They preserve the vector space P$(n)$.
The operators $q_a$ transform P$(n)$  into P$(n+2)$ and the operators
$\overline q_a(n)$ transform $P(n+2)$ into P$(n)$. 

The equations 
(\ref{caseN3}) provide a correspondence between the tensorial
operators (\ref{Up},\ref{Level},\ref{Down}) and linear
differential operators. All the identities
of section (3.4) are also obeyed by the differential operators.
Only the metric, say $\tilde g$, is different 
from our metric (\ref{gij1}) because the choice (\ref{caseN3})
corresponds to the normalisation~: 
$\tilde g_{+,-} = -2 , \tilde g_{0,0} = 1$. 

The operators preserving $P(n,n-2)$, which are at the root of
this work, correspond to
the representation $T_+$ for two levels and $s_t=n/2$.

The classification of linear differential operators preserving
P$(n_1, \ldots , n_k)$ \cite{bk} involves a number of generators
which  quickly grows with $N$. The generators close under an
appropriate 
choice for the commutators and anticommutators. The underlying algebraic
structure is, in this respect, still rather obscure.  
The sets of ten differential operators obtained by applying
the correspondence (\ref{caseN3}) to the representations constructed in the
previous sections allows one to write (considering the elements of their
envelopping algebra) the set of all differential operators underlying
the algebra ${\cal{A}}$ as a hidden symmetry.  

\section{Conclusions}

During the last years, many different algebras appeared to be relevant	
in several domains of theoretical physics~: graded algebras, Virasoro
and Kac-Moody algebras, W-algebras, $\ldots$. Some of these mathematical
structures can further be generalized and considered as deformed algebras
in the framework of quantum algebras.

The study of the hidden symmetries underlying the quasi exactly
solvable equations has revealed the occurence of yet other types
of graded (but not Lie) algebras, the ones called ${\cal{A}}(\Delta)$
in this paper. Given the integer $\Delta$, ${\cal A}(\Delta)$
contains $so(3)\times u(1)$ as a bosonic subalgebra and two sets
of fermionic generators, each of them transforming as a $spin$
$s=\Delta /2$ multiplet. In this respect, 
the ${\cal A}(\Delta)$ algebras extend 
the well know $N=2$ supersymmetric algebra $osp(2,2)$ 
with which it coincides
for $\Delta = 1$.

In this paper we have studied and classified the irreducible, 
finite dimensional representations of ${\cal{A}}(2)$. It appears
that the representations of this algebra possess a rich structure.
Namely, they assemble into five independant families plus one 
exceptional representation.  

Many new computations could be carried out in relation with the 
algebras ${\cal{A}}(\Delta)$ for $\Delta > 2$. For example~:
a concise formulation of their structure constants and the classification
of their representations.
More challenging is the construction of physical systems
admitting ${\cal{A}}(\Delta)$ as a hidden symmetry.
The interesting examples, known so far, are related to ${\cal{A}}(1)$
\cite{shifman,bk,bkgg,uly}. In absence of any real physical example
related to ${\cal{A}}(2)$, we simply mention a mathematical 
application which is discussed in \cite{bgk}~:
the finite dimensional representations of the Lie
superalgebra osp(3,2) can be formulated in terms of some
of the operators (\ref{caseN3}).

The algebraic structure ${\cal{A}}(\Delta)$ could also be looked at 
from the point of view of quantum deformations.
Indeed, considering finite difference QES equations (rather than differential
QES equations), it was recognized that the hidden algebra
becomes $sl(2)_q$, a deformation of $sl(2)$. Therefore, we can hope 
that some deformations of the algebra ${\cal{A}}(\Delta)$
will emerge from the study of finite difference QES systems.

\end{document}